# How Being Inside or Outside of Buildings Affects the Causal Relationship Between Weather and Pain Among People Living with Chronic Pain

Claire L. Little,[a] David M. Schultz,[b] Belay B. Yimer,[a] Anna L. Beukenhorst[c]

[a] *Centre for Epidemiology Versus Arthritis, Manchester Academic Health Science Centre, University of Manchester, Manchester, United Kingdom*

[b] *Centre for Atmospheric Science, Department of Earth and Environmental Sciences, University of Manchester, Manchester, United Kingdom*

[c] *Department of Biostatistics, Harvard T. H. Chan School of Public Health, Boston, Massachusetts*

File generated with AMS Word template 2.0

ABSTRACT

Although many people believe their pain fluctuates with weather conditions, both weather and pain may be associated with time spent outside. For example, pleasant weather may mean that people spend more time outside doing physical activity and exposed to the weather, leading to more (or less) pain, and poor weather or severe pain may keep people inside, sedentary, and not exposed to the weather. We conducted a smartphone study where participants with chronic pain reported daily pain severity, as well as time spent outside. We address the relationship between four weather variables (temperature, dewpoint temperature, pressure, and wind speed) and pain by proposing a three-step approach to untangle their effects: (i) propose a set of plausible directed acyclic graphs (also known as DAGs) that account for potential roles of time spent outside (e.g., collider, effect modifier, mediator), (ii) analyze the compatibility of the observed data with the assumed model, and (iii) identify the most plausible model by combining evidence from the observed data and domain-specific knowledge. We found that the data do not support time spent outside as a collider or mediator of the relationship between weather variables and pain. On the other hand, time spent outside modifies the effect between temperature and pain, as well as wind speed and pain, with the effect being absent on days that participants spent inside and present if they spent some or all of the day outside. Our results show the utility of using directed acyclic graphs for studying causal inference.

# SIGNIFICANCE STATEMENT

Three-quarters of people living with chronic pain believe that weather influences their pain. However, people staying inside would not be exposed to weather outside, and good weather may mean that people are more active outside, leading to more or less pain. To obtain data to calculate how the amount of time spent outside affects the weather–pain relationship, we conducted a 15-month smartphone study collecting daily pain reports and nearby weather for nearly 5000 participants in the United Kingdom. We found that time spent outside modifies the relation between temperature/wind speed and pain, showing the importance of accounting for other factors when investigating the association between weather and chronic pain, which could guide future research into pain mitigation and management.

# 1. Introduction

The belief that the weather affects the occurrence and intensity of pain in those people who live with chronic pain is quite old—going back to at least Hippocrates and ancient Chinese medical texts. Three-quarters of people who live with chronic pain believe that their pain levels are influenced by the weather (e.g., Hagglund et al. 1994; Timmermans et al. 2014). Despite this strong anecdotal evidence, however, scientific evidence to support their beliefs has not easily been forthcoming. A review of 43 previously published empirical studies that examined the weather–pain relationship found a wide range of results (Beukenhorst et al. 2020). For example, consider atmospheric pressure. Of the 38 studies reviewed by Beukenhorst et al. (2020) that investigated atmospheric pressure, 20 (53%) reported no link between pressure and pain, 11 (29%) reported high or increasing pressure was associated with higher pain levels, and 7 (18%) reported low or decreasing pressure was associated with higher pain levels. Other weather quantities such as temperature and relative humidity showed similar lack of consensus (Beukenhorst et al. 2020). This lack of scientific consensus leads many doctors to be skeptical about such a relationship, dismissing their patients' observations and concerns about the possible link between their pain and the weather and straining the doctor–patient relationship.

One of the principal reasons for this lack of consensus is the small sample size of these previous studies. Some studies used sample sizes as small as 20 individuals or collected data from the patients over just a few days (Beukenhorst et al. 2020, their Fig. 2). Given that other factors such as physical activity, sleep, and mood are more likely to be associated with people's pain than the weather and that weather can exhibit large changes on daily and seasonal time scales, large datasets are needed to tease out the subtle relationships between weather and pain. Modern digital epidemiological approaches such as using smartphone applications (i.e., apps) for more regular reporting and monitoring of people's pain may offer an alternative to traditional approaches in the past (e.g., Wilcox et al. 2012; Jardin et al. 2015; Solomon and Rudin 2020; Beukenhorst et al. 2022), although they are not without their challenges (e.g., De Montjoye et al. 2014; Bol et al. 2018; Mathews et al. 2019; Druce et al. 2019; Onnela 2020; Beukenhorst et al. 2022).

To collect the data to better understand the weather–pain relationship, the citizen-science project Cloudy with a Chance of Pain (http://www.cloudywithachanceofpain.com; hereafter, the *Cloudy project*) was created (e.g., Dixon et al. 2019). The Cloudy project was a United Kingdom–based smartphone app study to collect daily self-reported data on participants' pain. Each day, participants were prompted to answer ten questions about their pain and other measures of well-being (e.g., quality of sleep, mood, physical activity, time spent outside). For example, participants were asked "How severe was your pain today?" and reported pain on a 5-point scale: "no pain," "mild pain," "moderate pain," "severe pain," and "very severe pain." In addition, hourly measurements from the global positioning system (GPS) sensor in the phone (Beukenhorst et al. 2017) allowed the participant's location to be linked to the closest weather station in the Met Office observing network. In this way, we were able to develop individualized daily-averaged weather conditions that each participant experienced, accounting for their travels within the United Kingdom. Collecting these data allowed changes in pain to be linked to weather or changes in weather. Thus, the Cloudy project allowed for creation of a large dataset of participants (over 10,000 participants) with daily records of both pain and weather over a 15-month period.

Our research results from the Cloudy project show that, indeed, our large dataset can produce statistically robust relationships on the effects of weather on people living with chronic pain. Specifically, Dixon et al. (2019) used an epidemiological method called case-crossover to show that an increase in relative humidity, increase in wind speed, and decrease in pressure were associated with higher odds of a pain event in participants. Using synoptic compositing, Schultz et al. (2020) showed that the days with a high percentage of participants who experienced a pain event (defined here as a +1 change or greater in their pain levels on that 5-point scale) were associated with below-normal pressure, above-normal relative humidity, higher precipitation rate, and stronger wind. In contrast, days with a low percentage of participants who experienced a pain event were associated with above-normal pressure, below-normal relative humidity, lower precipitation rate, and weaker wind (Schultz et al. 2020). Yimer et al. (2022) found that between-participant variation existed, both in sensitivity to specific weather conditions and in the direction. Although there was no population-level association between temperature and pain, 1 in 10 participants with chronic pain were sensitive to temperature. Similarly, 1 in 25 were sensitive to relative humidity, 1 in 50 to pressure, and 3 in 100 to wind speed. The Cloudy project continues to yield opportunities for research on the influences on pain in participants.

In particular, one opportunity for deeper understanding of the weather–pain relationship is the self-reported data on the variables other than pain severity, which would allow testing whether the amount of time people spent outside influences their pain. In particular, two principal issues arise in using and interpreting this data in search of the answer to this question. First, being inside a building may be shelter from some weather conditions, but not others. For example, the temperature and humidity of air inside the building often differs from that outside,

File generated with AMS Word template 2.0

and being inside is protection from wind and precipitation. However, air pressure inside a building is generally in equilibrium with the air pressure outside. So, if the weather affects people's pain, such an effect may, for temperature and humidity, be contingent on whether people are outside. Second, weather affects the time people spent outside, as well as the amount of physical activity they do (e.g., Feinglass et al. 2011; Albrecht et al. 2020; Klimek et al. 2022). Specifically, in milder temperatures and in the absence of rain or snow, people spend more time outside and are more physically active. Physical activity may make pain worse or may make pain better, depending on the ailment and the individual. For example, most participants in our study were living with osteoarthritis or other musculoskeletal diseases (e.g., Dixon et al. 2019), and typically people with these conditions are recommended to do regular low-impact physical activity to reduce pain. Thus, all three variables—weather, time spent outside and physical activity—may influence pain.

Untangling the effects caused by these variables would help clarify the relationship between the weather, pain, and being inside/outside. The Cloudy project team collected information on the relative amount of time that people spent outside. Specifically, participants were asked "How much time have you spent outside today?" They reported back on a 5-point scale: "none," "some," "half," "most," and "all" of the day. Nonetheless, because participants recorded both their pain and time spent outside simultaneously, ascertaining whether pain affects a participant being outside or the weather affects a participant being outside, or both, is not easy to address. For this reason, in the analysis of the Cloudy project data by Dixon et al. (2019) and Schultz et al. (2020), time spent outside was not used as a variable in their analyses. Therefore, this present research aims to further understand the relationship between weather and pain by considering the potential causal relationships between weather, pain and time spent inside/outside.

This work is important because such improved understanding could lead to individualized forecasts of pain based on the weather forecast, giving those living with chronic pain the opportunity to self-manage their pain and its impacts on their own lives. Further understanding of the physiological processes that modulate pain could also lead to interventions to reduce pain and its impacts on individuals, as well.

The remainder of this article is organized as follows. In section 2, we propose models that explain the possible relationships between weather, pain, and time spent outside for four weather variables: air temperature, dewpoint temperature (a measure of moisture content in the air), atmospheric pressure, and wind speed. These models will be used to construct statistical tests to evaluate the strength of the relationships between time spent outside, weather, and pain, described in section 2. Section 3 describes the data and methods of the Cloudy project. Section 4 presents a sequential analysis testing each plausible relationship. Section 5 discusses limitations of the present study and offers future research opportunities. Section 6 summarizes the results and concludes the article.

## 2. Models of the possible ways that time spent outside may affect the causal relationship between weather and pain

Because the Cloudy project data recorded the time spent outside and the weather at the location of the participants, analysis of the results would seem to be straightforward. However, as discussed in section 1, there may be different ways to describe how time spent outside may affect the causal relationship between weather and pain. (Fig. 1). Here, we illustrate four of those causal relationships through the use of directed acyclic graphs (DAGs; e.g., Tennant et al. 2021).

1. Time spent outside is a *collider* of the weather–pain relationship (Fig. 1a). In this case, both weather and pain affected whether people spent time outside. This situation may be the case where the weather affected people's pain, which then determined whether they spent time outside, as well as the other route where the weather determined the amount of time spent outside.
2. Time spent outside is an *effect modifier* of the weather–pain relationship (Fig. 1b). In this case, the weather outside directly determined people's pain, but there was a different effect when time was spent outside. Therefore, the relationship between weather and pain was different among those participants and days when time was spent outside than when no time was spent outside.
3. Time spent outside is a *mediator* of the weather–pain relationship (Fig. 1c). In this case, weather affected whether people spent time outside (e.g., their exposure to the weather, their level of physical activity), which in turn influenced their pain. This situation is in addition to the direct relationship between weather and pain.
4. Time spent outside is a *confounder* of the weather–pain relationship (Fig. 1d). This confound would imply that the time spent outside influenced both the weather and people's pain, which would imply that the time spent outside by an individual influenced the weather. This implication is unrealistic, so we reject time spent outside as a confounder in the weather–pain relationship.

Thus, of these four possible models of the effects of time spent outside on the weather–pain relationship, the first three are all plausible. Yet, we do not know what model best explains the Cloudy project dataset. Therefore, we tackle the relationship between four weather variables (temperature, dewpoint temperature, pressure, and wind speed) and pain by proposing a three-step approach to untangle their effects. This approach follows that of Evans et al. (2012). First, we propose a set of plausible directed acyclic graphs that account for potential roles of time spent outside (e.g., collider, effect modifier, mediator), which we have done in Fig. 1. Second, we analyze the compatibility of the Cloudy data with the plausible models. Third, we examine the strengths of the resulting relationships and identify the most plausible model by combining evidence from the observed data and domain-specific knowledge.

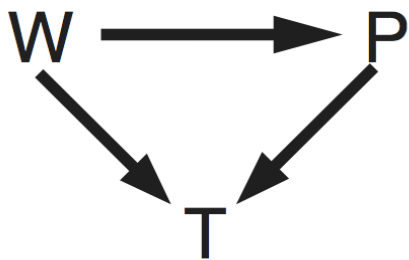
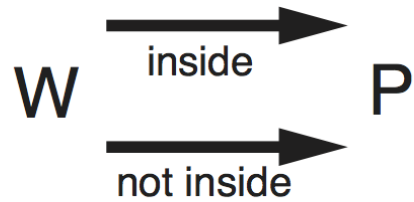

(a) collider

(b) effect modifier

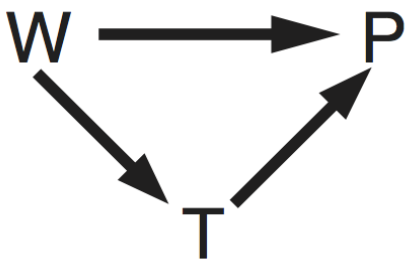
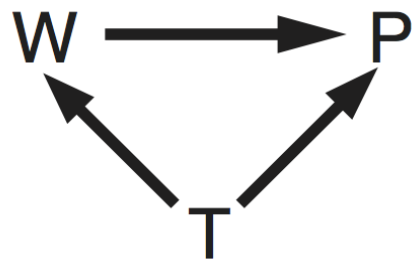

(c) mediator

(d) confounder

Fig. 1. Directed acyclic graphs of possible ways in which time spent outside (T) may affect the causal relationship between weather (W) and pain (P). (a) T is a collider of the weather–pain relationship, (b) T is an effect modifier of the weather–pain relationship, (c) T is a mediator of the weather–pain relationship, and (d) T is a confounder of the weather–pain relationship. Only (a), (b), and (c) are plausible relationships because time spent outside cannot influence the weather in (d).

## 3. Data

The description of the dataset largely derives from that of Schultz et al. (2020) as follows in the next five paragraphs.

The study ran for 15 months: 20 January 2016 to 19 April 2017. Before the study commenced, the app was tested and refined through interaction with a patient and public involvement group and a pilot study (Reade et al. 2017). Bolstered by two appearances on British Broadcasting Corporation (BBC) television and other national media attention (Druce et al. 2017; Fig. 2a in Dixon et al. 2019), a total of 13,207 users downloaded the study app during the 12-month recruitment period: 20 January 2016 to 20 January 2017. All 124 United Kingdom postcode areas were represented. Details about the demographics of the population can be found in the supplemental information in Dixon et al. (2019).

A total of 10,584 participants entered their demographic information and at least one pain report, making them eligible for the present study. A total of 6850 (65%) participants remained in the study beyond their first week and 4692 (44%) beyond their first month. Even after 200 days, 15% of participants were still entering data nearly every day (Druce et al. 2017; Fig. 2b in Dixon et al. 2019). This rate of engagement is exceptionally high compared to other mobile health studies where attrition often increases exponentially (e.g., Eysenbach 2005; Druce et al. 2017; Beukenhorst et al. 2022). We believe that our high retention is an indication of the easy-to-use app design, as well as the high level of interest by our participants in contributing toward an answer to this specific research question, which often has been of great personal interest to them (Reade et al. 2017; Druce et al. 2017, 2019).

To examine the relationship between time spent outside, weather, and pain for the Cloudy project dataset, we use the self-reported pain reports from the participants. How to use the pain reports, however, is not straightforward. The self-reported pain levels in our study could lead to bias by directly comparing one participant's pain levels to another. For example, two persons reporting "very severe pain" (5 on our 5-point scale) may have completely different experiences. In addition, participants may use the pain scale differently (e.g., never report maximum or minimum values of pain). To avoid this dilemma, we refer to studies [as reviewed by Olsen et al. (2017)] that show a 20% increase in pain is clinically significant. Such a clinically significant pain event could be measured by an increase of at least 1 category on our 5-point scale. As such, we define a *pain event* in an individual participant when they report a 1-category or greater increase in their pain level from the previous day (e.g., moderate pain yesterday to severe pain today).

To determine the importance of the relative amount of time spent outside to the weather–pain relationship, we separate the participants who provide a report into those who remain inside (those responding "none" to the question "How much time have you spent outside today?"; INSIDE) and those who spend at least part of the day outside (those responding "some", "half", "most", or "all of the day"; OUTSIDE).

As many as 17% of the participants entered only the demographic information and one day's report (Druce et al. 2017; Dixon et al. 2019). With no subsequent pain report on the next day, these very low engagers could never experience a pain event in our dataset. We also noticed higher levels of pain reported on days with spikes in recruitment, which was probably an artifact reflecting time needed for participants to settle into their personal scoring system or a "regression to the mean," where participants joined the study at times of higher than average pain but then settled into a more typical pattern. As such, we excluded data from all participants' first 10 days in the study.

The dataset used in this article is slightly different than that used in Schultz et al. (2020). Further quality control of the data has been performed, such as incomplete year-of-birth entries being excluded from further analysis. The number of respondents affected is small, resulting in small differences in the datasets that do not affect the results.

For each individual, a daily averaged set of weather conditions was calculated. The GPS sensor in the phone was used to determine an individual's location within the United Kingdom at each hour. In the case of location data being available only once a day, that information is used (Beukenhorst et al. 2017, 2021). That GPS location was linked to the closest weather station in the Met Office observing network. The weather data for the surface weather stations came from the National Oceanic and Atmospheric Administration/National Climatic Data Center Integrated Surface Database (https://www.ncdc.noaa.gov/isd). From these hourly weather reports, an individual daily averaged weather condition was associated with each person on each day. Four weather variables were used in this analysis: 2-m temperature, 2-m dewpoint temperature, sea level pressure (hereafter, pressure), and 10-m wind speed.

Statistics on the resulting dataset are as follows. There were 286,516 participant-days from 4963 participants that were used in this study. Of those, 62,796 were associated with pain events, and 223,720 were not associated with pain events. Of the pain-event participant-days, 13,852 were associated with the participant being classified as INSIDE, whereas 48,944 were classified as OUTSIDE. Of those participant-days not associated with a pain event, 46,962 were associated with the participant being classified as INSIDE, whereas 176,758 were classified as OUTSIDE. On any given day during the study, a minimum of 15.3% to a maximum of 28.8% of participants had a pain event from among the 130–1267 participants on any given day who entered their pain reports on two consecutive days. The mean rate of pain events for participants who would have entered daily data is 0.311, with a standard deviation of 0.191. Thus, the average person would have 8.7 pain events over a four-week period, if they entered data every day. By choosing "fraction of participants with a +1 increase in pain" as our outcome, our analysis is independent of how an individual interprets the pain scale and implicitly makes a within-person comparison.

## 4. Methods and results for each hypothesized causal relationship

In this section, we test the association between each weather variable and pain events under the assumptions of each of the three plausible models from section 2 using the statistical methods described below. For each model, we employ a mixed-effects logistic regression using the R package lmer. Logistic regressions are often used to fit to data with a binary outcome variable, which in the present study is a pain event. We accounted for the study design in that multiple pain measurements were collected from each participant over different days and that these repeated measures were correlated. Logististic regression for repeated measures can also increase precision of the effect estimates, as it accounts for unmeasured baseline differences between study participants.

To interpret mixed-effects logistic regressions, odds ratios are reported. For the present study, an odds ratio is the ratio of the probability of a pain event given a certain exposure to the probability of a pain event in the absence of that exposure (e.g., Szumilas 2010). Here, the exposures are the weather variables. Thus, an odds ratio greater than 1 suggests that the variable is associated with an increased risk of a pain event, whereas an odds ratio less than 1 suggests that the variable is associated with a decreased risk of a pain event. We also report on the confidence intervals at the 95% level. Statistical significance at the 5% level occurs when the confidence interval does not include 1.

*a. Is time spent outside a collider of the weather–pain relationship?*

If we assume that time spent outside is a collider of the weather–pain relationship (Fig. 1a), then both weather and pain affect an individual's time outside (i.e., $W \rightarrow T$ and $P \rightarrow T$). In this case, it would be erroneous to involve time spent outside in the model, as there is no causal relationship from time spent outside to another variable. In practice, if we assume a collider effect, this is equivalent to examining the relationship between weather and pain as if it were unaffected by time spent outside (i.e., W → P). In other words, what would the effect be if we ignored time spent outside and just looked at the relationship between the weather and pain?

To model this relationship under the assumption of a collider, we employ mixed-effect logistic regressions for each weather variable in turn. Each mixed-effect logistic regression is fitted with pain events as the outcome variable, the weather variable as an explanatory variable, and a subject-specific random intercept term. Coefficients of the explanatory variables are calculated by the model. We exponentiate these coefficients to provide an odds ratio for each weather variable.

The results of the logistic regression show that, pressure, and wind speed have a small statistically significant relationship with pain events, but that dewpoint and temperature do not have a statistically significant relationship with pain events (Table 1). Specifically, an increase in pressure by 1 hPa reduces the odds of a pain event by 0.3%, and an increase in wind speed by 1 m s$^{-1}$ increases the odds of a pain event by 1% (Table 1). For larger changes in the weather, an increase in pressure by 10 hPa reduces the odds of a pain event by 2.5%, and an increase in wind speed by 10 m s$^{-1}$ increases the odds of a pain event by 6.1% (not shown). Although these effects are statistically significant, their small magnitudes mean that these effects may not be clinically meaningful.

Table 1: Effect of the four weather variables on pain in univariable models by not correcting for time spent outside (i.e., assuming time spent outside is a collider), including the lower and upper bounds on the 95% confidence interval around the odds ratio.

| Model | Odds ratio (95% confidence interval) |
|---|---|
| **Temperature** | 0.998 (0.996, 1.000) |
| **Dewpoint temperature** | 0.999 (0.998, 1.002) |
| **Pressure** | 0.998 (0.997, 0.998) |
| **Wind speed** | 1.010 (1.005, 1.015) |

If time spent outside is a collider, then collider bias would be introduced by erroneously including time spent outside in the model. This collider bias is the departure of the conditional association of the weather and pain events from their marginal association. We assess this collider bias by examining the difference between the coefficients in the above model (i.e., W → P) to those of an alternative model that includes time spent outside as an explanatory variable (i.e., W → T). When compared with the odds ratio of the alternative logistic regression that includes time spent outside as an explanatory variable (Table 1), pressure and wind speed have a small statistically significant association with the odds of a pain event, whereas temperature and dewpoint temperature do not have significant associations. Specifically, the odds ratio for temperature declines by 0.0003, the odds ratio for dewpoint temperature declines by 0.00019, the odds ratio for pressure declines by 0.00007, and the odds ratio for wind speed increases by 0.0004. These changes in odds ratios are lower than 0.05%. Thus, if we incorrectly induced collider bias by adjusting for time spent outside, it would have minimal impact on our findings.

To summarize, two of the four variables (i.e., pressure, wind speed) show statistically significant relationships with pain events, but these are small effects. If time spent outside were a collider, a model that corrects for that variable would change the relationship, albeit not by much. Thus, we conclude that the collider effect is not a strong effect on an already weak weather–pain relationship.

*b. Is time spent outside an effect modifier of the weather–pain relationship?*

An effect modifier (Fig. 1b) is equivalent to saying that there is a difference between the relationship between weather and pain events on the days that people spend inside compared to days people spend outside, but that neither the weather nor the pain event would be the cause for people to stay inside in the first place. We can test for the effect modifier by including an interaction term between weather and time spent outside. In practice, because INSIDE/OUTSIDE is a binary variable, the effect modifier is equivalent to examining two separate W → P relationships, with one being for when the participants are inside (i.e., those responding "none" to the amount of time spent outside) and one for being when the participants are not inside (i.e., those responding "some", "half", "most", or "all of the day"). If these two relationships differ, then the time spent outside is an effect modifier. If these two relationships are the same, then the time spent outside is not an event modifier.

To examine time spent outside as an effect modifier, we employ mixed-effects logistic regression models for each weather variable. In this case, pain event is the outcome variable and there are three explanatory variables: the weather variable, the INSIDE/OUTSIDE variable, and an interaction term between the two. The coefficient of the weather variable is analyzed for those participant-days that were spent entirely inside. The coefficients of all three explanatory variables are summed for those participant-days where some time was spent outside. In both cases, the combination of coefficients is exponentiated to calculate the odds ratio of a pain event.

If we assume that time spent outside is an effect modifier, then the results of this analysis show the following (Table 2).

There is no association between dewpoint temperature and the odds of a pain event, neither when all time is spent inside nor when some time is spent outside. Dewpoint temperature is a measure of humidity. Although outside humidity drives humidity inside buildings (i.e., if it is more humid outside, it is typically also more humid inside buildings), buildings provide some protection from outside humidity. Inside humidity is further influenced by activities in the building; through breathing, cooking, running water taps, and boiling water, indoor humidity may increase, and through air conditioning or dehumidifiers it may decrease. People who spend a lot of time inside, may therefore be exposed to a different ambient humidity—any effect of inside humidity rather than outside humidity cannot be excluded based on this analysis.

For pressure, the odds of a pain event are the same for days spent inside and days spent outside. There is no statistically significant difference depending on time spent outside. This result can be explained by pressure being the same inside and outside buildings.

For wind speed, the odds of a pain event are the same for days spent inside and days spent outside. However, the association between wind speed and pain events is not statistically significant for days spent inside, whereas it is for days outside. Hence, the association between wind speed and pain events may be absent if all time is spent inside, whereas higher wind speed increases the odds of a pain event slightly if some time is spent outside. This result may be explained by people being exposed to wind speed outside buildings, but not when they are inside buildings. At higher temperatures, the odds of a pain event are lower, but only on days spent outside. On days spent inside, there is no statistically significant association between temperature and pain events. (Table 2).

Thus, time spent outside is not an effect modifier for dewpoint temperature or pressure. Time spent outside is an effect modifier of the relationship between wind speed and pain and between temperature and pain. For both wind speed and temperature, the association is significant when time is spent outside, but not when all time is spent inside. When people spend time outside, higher temperatures and lower wind speeds are associated with lower odds of a pain event. For wind speed, the basic model without effect modification already shows an association with pain events (even though it sums effects for days spent inside and outside).

For temperature, the basic model without effect modification does not show an association with pain events. Accounting for effect modification is needed to uncover the association between temperature and pain events on days with time spent outside.

Table 2: Results of univariable models considering time spent outside as an effect modifier. Estimates with 95% confidence intervals in parentheses.

| Weather variable | Odds ratio for INSIDE (95% confidence interval) | Odds ratio for OUTSIDE (95% confidence interval) |
|---|---|---|
| **Temperature** | 1.000 (1.000, 1.010) | 0.997 (0.995, 0.999) |
| **Dewpoint temperature** | 1.000 (1.000, 1.010) | 0.999 (0.997, 1.001) |
| **Pressure** | 0.997 (0.995, 0.999) | 0.998 (0.997, 0.999) |
| **Wind speed** | 1.010 (1.000, 1.020) | 1.010 (1.004, 1.015) |

*c. Is time spent outside a mediator of the weather–pain relationship?*

The mediator effect is explained by comparing the W $\to$ P relationship to the W $\to$ T $\to$ P relationship (Fig. 1c). Hence, testing the mediator effect only makes sense for the weather variables that have a statistically significant relationship with pain (Baron and Kenny 1986). In the mixed-effect logistic regressions, pressure and wind speed are statistically significant with the time spent inside ($p < 10^{-16}$ in each case; not shown). Dewpoint temperature and temperature do not have a statistically significant relationship with pain (Table 1) and are therefore excluded from this analysis.

For each model in Table 3, we report the average direct effect (ADE, the direct effect of time spent outside on pain events, W $\to$ P), the average causal mediation effects (ACME, the indirect effect of the weather variable on the pain event that goes through time spent outside, W $\to$ T $\to$ P), and the total effects (ADE + ACME). We then use these quantities to calculate the *percentage mediated* (i.e., the percentage of the weather–pain relationship attributed to time spent outside = ACME × 100% ÷ (ADE + ACME)).

The results from these analyses can be interpreted as follows (Table 3). For pressure, the indirect effect is not significant, and the proportion mediated is not significant and small (much less than 1%).

For wind speed, the indirect effect, direct effect, and overall effect are all statistically significant. The proportion of the W $\to$ P relationship that is mediated by the W $\to$ T $\to$ P relationship is 4.36% (95% CI: 2.32–9.07%), with an indirect effect size of 0.0000699 (Table 3). Hence, on days with higher wind speed, participants are more likely to stay inside, explaining a small part of the increase in pain events on these days. Of note, the magnitude of the indirect effect is much smaller than the direct effect. It is unlikely to be clinically meaningful: a one unit increase in wind speed results in a 0.0000699 increase in the probability of a painful event because of spending more time inside following the higher wind speed. Although the mediation effect is statistically significant, it is not a clinically important driver of pain.

Table 3: Model results regarding time spent outside as a mediator effect of the relationship between the weather variable and pain, with 95% confidence intervals in parentheses, for (a) pressure, and (b) wind speed. The average direct (ADE), indirect effects (ACME), and the total effect (ADE+ACME) are presented in risk difference scale in line with the mediation R package.

| (a) Model | Effect | Odds Ratio (95% confidence interval) |
| --- | --- | --- |
| **Mediator model** | Pressure | 0.991 (0.990, 0.992) |
| **Outcome model** | Pressure | 0.998 (0.997, 0.998) |
|  | Inside | 1.049 (1.025, 1.074) |
| **Mediation** | ADE | –0.000420 (–0.000434, –0.000357) |
|  | ACME | –0.000000135 (–0.00000203, –0.00000156) |
|  | ADE + ACME | –0.000419 (–0.000435, –0.000357) |
|  | % Mediated | 0.00115% |

| (b) Model | Effect | Odds Ratio (95% confidence interval) |
| --- | --- | --- |
| **Mediator model** | Wind speed | 1.071 (1.066, 1.077) |
| **Outcome model** | Wind speed | 1.010 (1.005, 1.014) |
|  | Inside | 1.049 (1.025, 1.074) |
| **Mediation** | ADE | 0.00156 (0.000845, 0.00232) |
|  | ACME | 0.0000699 (0.000391, 0.000104) |
|  | ADE + ACME | 0.00163 (0.000917, 0.00238) |
|  | % Mediated | 4.3% |

# 5. Limitations of the present study and future research opportunities

There are a number of limitations to our present study, and these provide possible future research opportunities.

1. Our results were obtained from a study in the United Kingdom, a midlatitude location downstream from the North Atlantic Ocean. As different people may have acclimatized differently in different climates, it is difficult to understand how these results could be generalized beyond the United Kingdom. Therefore, there is the opportunity to explore the possible effect on pain of weather, as well as other possible effects, in other locations around the world.

2. Our daily collection of pain and well-being information from participants sets the time scale for which we can assess the weather–pain relationship. This time scale was set by the anecdotal evidence that patients generally report pain and weather fluctuating on a time scale on the scale of days. Within-day collection of this information would allow study of sub-daily effects, but would also require more frequent reporting of pain and well-being indicators by the patients, which would be tedious and risk reducing their participation. Furthermore, a greater range of activities would need to be included in the well-being indicators, as well.

3. Untangling the weather–pain relationship is difficult, and our study focused on a subset of possible relationships illustrated through directed acyclic graphs that capture the interplay between weather, time spent outside, and pain. However, there may be other potential directed acyclic graphs involving additional nodes or feedback loops that might further elucidate the relationship. For example, one might consider the application of discovery algorithms to explore other directed acyclic graphs (e.g., Ramsey et al. 2017; Li et al. 2020), but these methods have their own limitations. One might extend the directed acyclic graphs with variables that may confound any relationship between time spent outside and pain, such as physical activity.

4. Greater precision might be needed in locating the weather conditions experienced by the participants. For example, many weather stations may be located at airports and other locations that are not representative of where people live. Urban heat island, mountain/valley, coastal, and other microclimatic effects would produce noise in any actual signal of the weather–pain relationship. Such challenges could be overcome by having people carry portable weather sensors with them. However, such efforts would again limit participation in future studies.

5. In the present study, we examined the population-wide effect. However, there might be variability among the population by disease or among certain weather-sensitive individuals (e.g., Yimer et al. 2022). Therefore, future research might investigate these subpopulations.

6. This study did not measure sunshine because the Met Office stations used do not collect that data. But, a sunny day with the otherwise exact same weather condition to a cloudy day will likely encourage more people to go outside. Future studies should investigate the amount of sunshine and its effect on the weather–pain relationship.

7. Other exposure variables could be considered, such as whether the inside location is heated or air conditioned relative to the outside. In addition, body-worn portable technology may enable studies that measure ambient temperature, humidity, and pressure, further reducing measurement error.

8. Other measures of comfort could be considered, such as the human comfort index, wind-chill temperature, and other heat- and cold-related indices.

## 6. Conclusions

Although the belief that the weather affects one's pain has long been held by many people, scientific evidence in support of a consensus has been difficult to achieve, in part because of small data sample sizes required to tease out subtle relationships (e.g., Beukenhorst et al. 2020). These limitations can be overcome through smartphone citizen-science experiments over an extended period of time, as was achieved in Cloudy with a Chance of Pain, which involved over 10,000 participants for a 15-month period (e.g., Dixon et al. 2019). When such limitations are overcome, the popular belief that the weather influences pain events can be confirmed (e.g., Dixon et al. 2019; Schultz et al. 2020; Yimer et al. 2022).

Further analysis of the Cloudy project dataset also shows the value of collecting patient-reported outcomes that help untangle some of the relationships between different weather variables and the time people spend outside. For example, being inside a building is shelter from temperature, dewpoint temperature, and wind, but not pressure. Another example is the causation direction between weather, pain, and time spent outside. If low pain levels caused by favorable weather result in people spending more time outside being physically active and ending up with pain events, then weather is not a simple factor affecting pain only. To understand the possible ways in which time spent outside may affect the causal relationship between weather and pain, we posit four possible relationships for time spent outside: collider, effect modifier, mediator, and confounder. We rule out time spent outside as a confounder because it would imply that the time spent outside by a participant influences the weather. Further, our analysis showed that time spent outside as a collider or mediator of the weather–pain relationship is not supported by the data. Time spent outside is not an effect modifier for the relationship between dewpoint temperature or pressure and pain events. Instead, we found evidence that the effect of temperature and wind speed on pain events are modified by time spent outside: the effect is present on days that time is spent outside, but is not present when a participant was inside all day.

If our results are valid and physiological processes can be identified that modulate pain, our results will be important for understanding how patients can learn to deal with chronic pain and its association with weather. These results would suggest approaches for mitigation of pain events in sensitive populations. Specifically, for people whose pain is sensitive to temperature, although the weather has a direct effect on pain, it also influences pain through time spent outside. Thus, by being outside, these individuals may manipulate or counteract part of their pain from the direct effect. In contrast, for people whose pain is sensitive to pressure, there is little difference between being inside or outside.

*Acknowledgments.* We thank the 10,584 participants in Cloudy with a Chance of Pain who made this study possible with their dedication to daily reporting of their pain levels. We thank Will Dixon for leading Cloudy with a Chance of Pain. We are grateful for the contributions of our patient and public involvement group throughout the study: Carolyn Gamble, Karen Staniland, Shanali Perara, Simon Stones, Rebecca Parris, Annmarie Lewis, Dorothy Slater, and Susan Moore. We thank Bruce Hellman and Ben James at uMotif Limited, London, for their assistance in app design, data collection, and storage. The study app and website were provided by uMotif Limited. The unique flower-like "motif" symptom-tracking interface is owned by uMotif Limited and protected through EU Design Registrations and a US Design Patent. We thank John McBeth, Huai Leng Pisaniello, Thomas House, Indrani Roy, and two anonymous reviewers for their input and comments on earlier drafts of this article. Cloudy with a Chance of Pain was funded by Versus Arthritis (grant reference 21225), with additional support from the Centre for Epidemiology (grants 21755 and 20380). Beukenhorst was supported by a UK Medical Research Council doctoral training partnership (grant MR/N013751/1). Schultz was partially supported by the Natural Environment Research Council UK (grants NE/I005234/1, NE/I026545/1, NE/N003918/1 and NE/W000997/1). Schultz performed part of this work at the Aspen Center for Physics, which is supported by National Science Foundation grant PHY-2210452.

*Data Availability Statement.* The National Oceanic and Atmospheric Administration/National Climatic Data Center Integrated Surface Database (https://www.ncdc.noaa.gov/isd) provided the weather data from surface stations. Access to the Cloudy with a Chance of Pain dataset is not available at this time.

File generated with AMS Word template 2.0